# Deciphering quantum fingerprints in electric conductance

Shunsuke Daimon[1,2,†,*], Kakeru Tsunekawa[1,†], Shinji Kawakami[1], Takashi Kikkawa[1,3,4], Rafael Ramos[3,‡], Koichi Oyanagi[4,5], Tomi Ohtsuki[6] & Eiji Saitoh[1,2,3,4]

[1]*Department of Applied Physics, The University of Tokyo, Tokyo 113-8656, Japan.*

[2]*Institute for AI and Beyond, The University of Tokyo, Tokyo 113-8656, Japan.*

[3]*WPI Advanced Institute for Materials Research, Tohoku University, Sendai 980-8577, Japan.*

[4]*Institute for Materials Research, Tohoku University, Sendai 980-8577, Japan.*

[5]*Faculty of Science and Engineering, Iwate University, Morioka 020-8551, Japan.*

[6]*Physics Division, Sophia University, Chiyoda, Tokyo 102-8554, Japan.*

[†]There authors contributed equally: Shunsuke Daimon, Kakeru Tsunekawa

[‡]Present address: Centro de Investigación en Química Biolóxica e Materiais Moleculares (CIQUS), Departamento de Química-Física, Universidade de Santiago de Compostela, Santiago de Compostela 15782, Spain

[*]Correspondence and requests for materials should be addressed to S.D. (email: daimon@ap.t.u-tokyo.ac.jp).

**Abstract**

**When the electric conductance of a nano-sized metal is measured at low temperatures, it often exhibits complex but reproducible patterns as a function of external magnetic fields, called quantum fingerprints in electric conductance. Such complex patterns are due to quantum-mechanical interference of conduction electrons; when thermal disturbance is feeble and coherence of the electrons extends all over the sample, the quantum interference pattern reflects microscopic structures, such as crystalline defects and the shape of the sample, giving rise to complicated interference. Although the interference pattern carries such microscopic information, it looks so random that it has not been analysed. Here we show that machine learning allows us to decipher quantum fingerprints; fingerprint patterns in magneto-conductance are shown to be transcribed into spatial images of electron wave function intensities (WIs) in a sample by using generative machine learning. The output WIs reveal**

**quantum interference states of conduction electrons, as well as sample shapes. The present result augments the human ability to identify quantum states, and it should allow microscopy of quantum nanostructures in materials by making use of quantum fingerprints.**

**Introduction**

In metals at low temperatures, the quantum-mechanical wave nature of conduction electrons comes to the fore, which can be described in terms of their wave functions. The phase of the wave function can be modulated by a magnetic field, causing wave interference of electrons as a function of external fields[1,2]. The simplest case can be found in a ring sample where the wave propagation is completely restricted along the circumference. In such a sample, an electron circling along the ring gains a phase proportional to the magnetic field inside the ring. If the phase is an even multiple of $\pi$, constructive interference of the electrons enhances the wave nature and electric conductance. When the phase is an odd multiple of $\pi$, on the other hand, destructive interference reduces electric conductance, giving rise to a periodic oscillation of the conductance with respect to the external field, called an Aharonov-Bohm oscillation[3,4,5,6]. In a real nanometal, however, there are many scatterers for the electron waves, such as impurities and nanostructures (Fig. 1a, b). As a result, interference among all electron waves propagating among the scatterers piles up to modulate the conductance, giving birth to a complicated magnetic field dependence reflecting the distribution of the scatterers[7,8] (Fig. 1c). These complex patterns of electric conductance are called conductance fluctuations or quantum fingerprints in electric conductance[1]. The fingerprint thus carries information concerning the quantum electron states. Nevertheless, it has been considered difficult to interpret it due to its complexity. In the literature (refs. 9-11), conductance was shown to be predicted from microscope information such as defect positions by using machine learning methods. Here we show that, by developing machine learning[12,13] for quantum interference, fingerprint patterns in magneto-conductance can be transcribed into a real image of electron WIs and a sample shape.

Feature extraction and geometry generative deep-neural networks are combined to reconstruct electron WIs and sample shape images from the magneto-conductance data (Fig. 1d). We named the present network a quantum geometric decoder (QGD). As a training dataset, we use numerical-calculated conductance and WIs in two-dimensional nanowires with antidot defects exposed to magnetic fields (Fig. 2). The network (A) + (B) in the QGD (Fig. 1d) extracts essential features of the WI images using a dimensionality-reduction technique based on a variational autoencoder[14,15] (VAE) (Fig. 3a). By training the network to compress the calculated real-space WI data onto a low dimensional latent space and to reconstruct the data from the latent-space data, the network (A) learns to convert the WI images into a vector in the latent space so as to best reproduce the original images using the network (B), and the network (B) to generate WI images from the compressed information represented by the vector (Fig. 3a). In the following, we will show that, owing to the latent-space information, the QGD can find relations between complicated quantum interference and quantum fingerprints, which cannot be realised by conventional methods.

**Results**

**Calculation of wave function intensities and conductance**

To prepare a training dataset, we perform numerical calculation; Fig. 2a shows a calculation model of the two-dimensional nanowire with a size of $60 \times 50$ single-orbital sites. For simplicity, defects in the nanowire are introduced as two antidots with a radius of 5 sites. One antidot is fixed around the upper centre of the system. With 1591 different antidot positions and 5 different random potentials applied to the systems, 7955 samples are prepared as a dataset. For each sample, we calculate WIs and two-terminal magneto-conductance by using a tight-binding method and Landauer's formula (see Methods), where the magnetic field is introduced as a Peierls phase in the Landau gauge and inelastic scattering is not taken into consideration, assuming extremely low temperatures.

Figure 2c exemplifies the calculated magneto-conductance, $G$, for a sample as a function of the externally applied magnetic flux density $B$. Each conductance data exhibits different complex patterns. To highlight the patterns, we introduce the normalised magneto-conductance $\Delta G(B)$ calculated by subtracting the averaged $G$ over all the 7955 conductance data from the raw $G(B)$ data (see Fig. 2d). We also calculate the wave functions under the zero magnetic flux density condition. We then obtain the WI images by squaring the modulus of the wave functions. Figure 2b shows WI images, where complicated quantum interference can be seen around the antidot defects. Zero padding[16] with a width of 5 pixels is added to the left and right ends of each $60 \times 50$ pixel WI image, resulting in the final size of $60 \times 60$ pixels.

**Quantum geometric decoder network**

The QGD is trained to generate correct sample-shape images and quantum interference patterns from $\Delta G(B)$ (Fig. 1d). To realise the generation, we combine a feature extraction network [(A) + (B) in Fig. 1d] and a geometry generative network [(C) + (B) in Fig. 1d] in the QGD. The feature extraction neural network is based on VAE[14,15]; VAE was shown to compress its input and extract the information necessary to reproduce the input on the output of the network[14]. The network converts a WI image with 3600 pixels into a 7-dimensional latent-space data [(A) in Fig. 1d] and then converts the data back to an image with 3600 pixels [(B) in Fig. 1d]. The network is trained so that the output image matches the input image (see Supplementary Fig. 1a, Tables 1, and 2 for more details). After the training, the latent space of the QGD acquires essential information to reconstruct the WI images.

The QGD can directly generate WI images from magneto-conductance data $G(B)$ (Fig. 4a and b). First, the network (C) in Fig. 1d connects the input 101-points $\Delta G(B)$ data and the 7-dimensional latent space, and then generates a WI image using the deconvolution part (B) of the VAE-based network. We trained the fully-connected neural network (C) so that the generated image well reproduces the input WI image associated with the magneto-conductance data (see Supplementary Fig. 1b and Table 3 for more details). Such a Y-shaped QGD was found to acquire an excellent capacity for WI generation as follows.

**Deciphering quantum fingerprint in magneto-conductance**

Figure 4a, b shows a typical result of deciphering a quantum fingerprint into a WI image by using the QGD. Surprisingly, the QGD spontaneously generates a clear WI image just from conductance data (see Fig. 4a, b). The generated WI image and sample shape coincide with a separately calculated WI image (Fig. 4d) and the corresponding sample shape (Fig. 4c), respectively. To evaluate the generation fidelity, we calculated the root mean squared (RMS) error between the generated and calculated images for each sample in the test dataset (see Methods for more details). The average RMS error is $2.1 \times 10^{-5}$. More examples are shown in Fig. 4e-g; WI images are almost correctly generated for all the magneto-conductance inputs. In addition to the locations of the antidots, significantly, the quantum interference patterns of the WIs are well generated from only the magneto-conductance data (see the fringe patterns in Fig. 4f). We also checked that similar generalisation performance of the QGD can be obtained even if some different system parameters are used (see Methods). The results show that, although a quantum fingerprint pattern looks random, it contains information on the quantum interference in the sample, and the QGD can interpret it.

**Discussion**

To check the performance of the feature extraction network [(A) + (B) in Fig. 1d], in Fig. 3a, we show the result of the WI autoencoding by the trained network. The output image clearly reproduces the input WI image (see Supplementary Figs. 3 and 10 for detailed error profiles), which can be attributed to the fact that the sharpness of the image generation is an advantage of the VAE. The RMS error is $2.0 \times 10^{-5}$, comparable to the error of the QGD output. More examples are shown in Supplementary Fig. 3a-c. It is surprising that such complicated geometry information is reconstructed using only the 7-dimensional data in the latent space, which is three orders of magnitude smaller than the dimension of the input data: 3600. This demonstrates the excellent compressing ability of the feature extraction network.

The above decipherment using the QGD means that the present network can analyse the quantum fingerprints to reconstruct the microscopic images (geometry and WIs) in the sample. In the network, information on the interference is compressed into the 7-dimensional latent space. Each WI image is convoluted into a Gaussian distribution in the latent space, and the set of the distribution forms a global geometric structure describing similarity among the states. We now visualise and discuss the geometric structure of the data in the latent space. To this end, we use a dimensionality reduction method called the universal manifold approximation and projection (UMAP)[17], which converts data points in the 7-dimensional latent space into those in a 3-dimensional space while keeping their topological structure. Figure 3c shows the obtained 3-dimensional representation of the latent-space data. Owing to the variational Bayesian approach of VAE[14], VAE was found to construct well-organised data structures in the latent space (see also Supplementary Fig. 12). In the present feature extraction network, one can see that the dataset forms a 2-dimensional curved surface in the latent space (Fig. 3c), but the surface shows a structure with some thickness (see, for instance, the regions with a dual-layered structure indicated by the red arrows in Fig. 3c). The 2-dimensional nature of the dataset is consistent with the degree of freedom of the antidot location $(x, y)$. However, the appearance of the data-point scattering along the thickness direction cannot be explained by the location degree of freedom.

In order to understand the geometric structure, we performed a control experiment using geometry images without WIs (compare Fig. 3a, c and b, d and see Methods for details of the geometry images without WIs). In contrast to the images which contain WI as well as sample-shape images, the dataset without WI images forms a simple plane without the thickness structures in the 3-dimensional representation (see Fig. 3d). By comparing the data in the latent space, we found that the dataset generated from the inputs with WI images is about ten times thicker than that generated from the inputs without WI images, where the thickness is defined as the variance of the data points (see Methods and Supplementary Fig. 11). The result suggests that the thickness structures, such as the dual-layered structure, carry electron information represented by WIs. The dual-layered structure was found to be related clearly to the antidot location and interference; the upper layer consists of the data points for even $x$, while the lower one for odd $x$ in the present parametrisation [see the points coloured in yellow (for even $x$) and blue (for odd $x$) around the red arrows in Fig. 3c]. To show how the interference patterns change with the antidot location $(x, y)$, we calculate the absolute difference and the RMS difference of the WI images between $(x, y)$ and $(x + \Delta x, y)$. Figure 3f, g shows the results at $(x, y) = (12, 33)$ for $\Delta x = 1, 2$, respectively. The absolute difference for $\Delta x = 2$ is less than that for $\Delta x = 1$ [compare the images between Fig. 3f and g, *e.g.*, at the left and right sides of the antidot position $(x, y)$]. In Fig. 3h, we plot the RMS difference in each image as a function of $x$ at $y = 33$. For most of the $x$ values, the RMS difference for $\Delta x = 2$ turned out to be less than that for $\Delta x = 1$, indicating that the WI image for the antidot position $(x, y)$ has similar features to that for $(x + \Delta x, y)$ with $\Delta x = \pm 2, \pm 4, \pm 6 \cdots$ in the present parametrisation. In fact, we found that, near the antidot, the crest and trough patterns of the WIs agree with each other between $(x, y)$ and $(x + \Delta x, y)$ for even $\Delta x$ numbers, while those are reversed for odd $\Delta x$, suggesting that, in the present study, the constructive and deconstructive interference gives rise to the observed dual-layered structure in the latent space (see Methods for more details). The result suggests that the electron information, such as interference, is encoded in the latent space as the thickness-dimension information, which emerges on the 2-dimensional curved surface representing the antidot location. Such extra-dimension information appears to allow the QGD to interpret quantum fingerprint patterns in the present study.

In summary, we demonstrated the decipherment of quantum fingerprints in conductance by developing QGD. The QGD was found to transcribe the complicated magneto-conductance patterns into geometric information such as defect distributions and WIs in the samples. The WI patterns turned out to be encoded in the latent space of QGD as an extra dimension of the manifold representing the defect position information. It is truly worthwhile to tune the network by using data from real objects to show the versatility of the present method (for experiments using physical samples, see Supplementary Fig. 4 and Supplementary Note 1). We expect that a wide range of signals from quantum systems can be interpreted by this method.

**Methods**

**Numerical calculations of the input dataset.** The WIs and two-terminal magneto-conductance calculations are based on a tight-binding model and the Landauer-Buttiker formula[2]. The Hamiltonian for the present calculation is $H = \sum_i (4t + U_i) c_i^\dagger c_i - \sum_{<i,j>} t \exp[-i \pi B a^2 / \phi_0 (x_i - x_j)(y_i + y_j)] c_j^\dagger c_i$, where $c_i^\dagger$ and $c_i$ are creation and annihilation operators, respectively, for electrons at the site $(x_i, y_i)$, $t$ is the hopping energy, $U_i$ is a random potential representing sample-specific randomness, and $\phi_0 \equiv h/e$ is the magnetic flux quantum. The magnetic flux density $B$ is introduced over the entire scattering region. $a$ is the lattice constant and it is not a fixed parameter since our model is spatially scale-free calculation[18]. The second term in the Hamiltonian is a Peierls phase with the Landau gauge and is summed over the nearest-neighbor pairs. The hopping energy is set to $t = 1$. The magnitude of the disorder potential is set to one tenth of the hopping energy; $U_i$ is sampled uniformly but randomly in the range from -0.05$t$ to 0.05$t$. The Fermi energy is set to 2.0$t$ except for the dataset of the Fermi-energy dependent magneto-conductance calculation in Supplementary Fig. 8. One antidot is fixed at the upper centre of the system, while the other is located so as not to overlap with the fixed one. The antidot is modelled by removing lattice points in a circular shape from the 2-dimensional square lattice of the nanowire. Two leads with the same square lattice and hopping energy are attached to both ends of the scattering region. Kwant[18], a code for numerical calculation of wave functions and quantum transport properties, is used to obtain the dataset of magneto-conductance and WIs for the samples with antidot defects. $B$ dependence of the 2-terminal conductance is calculated, where $B$ is swept from 0 to 0.12 in 101 divisions. The normalised magneto-conductance $\Delta G(B)$ is calculated by subtracting the averaged conductance $G_{\text{ave}}(B)$ from the raw $G(B)$, where we defined $G_{\text{ave}}(B)$ as the conductance value averaged over 7955 conductance data with 1591 different antidot positions and 5 different random potentials. We checked that the conductance calculation works well (see Supplementary Figs. 13 and 14) and $\Delta G(B)$ shows fluctuation with respect to $B$ with the variance comparable to $e^2/h$[19]. WIs are calculated at the zero magnetic flux density condition. We note that a part of wave function phase information is included in the WI because the amplitude is the result of complex interference between scattered waves. The WI images shown in the figures are normalised such that the sum of the intensity equals to unity for each antidot configuration; $\sum_{i=1}^{60} \sum_{j=1}^{60} X_{i,j} = 1$, where **X** is a WI image, and the suffixes $i, j$ represent the pixel index. Note that the normalisation is done only for the visualisation plots after the training, and unnormalised data is used during the training. A geometry image without WIs is defined as an image with a non-zero constant value on a nanowire and zero inside the antidots and outside the nanowire. No WI images are superimposed. The constant value is determined by the normalisation condition $\sum_{i=1}^{60}\sum_{j=1}^{60} X_{i,j} = 1$, where **X** is an image, and the suffixes $i,j$ represent the pixel label.

**Feature extraction network and its training.** The feature extraction network is based on a VAE network[14,15] comprising image encoding and decoding networks. The encoding network is composed of 2 two-dimensional convolution layers with the kernel size of $4 \times 4$ and 3 fully-connected layers, where each layer has 1024, 512 and 7 nodes. The encoding network outputs a 7-dimensional gaussian distribution in the latent space via the reparameterisation trick used in VAE[14,15]. The dimension of the latent space is set to 7 to reconstruct the input WI images with high accuracy. The decoding network is composed of 2 fully-connected layers and 2 two-dimensional deconvolution layers with the kernel size of $4 \times 4$. The network architecture parameters are determined based on a recipe found in machine learning for image recognition[15] and then fine-tuned. Rectified linear units (ReLUs)[20] and Leaky ReLUs[21] are used as activation functions. To improve the learning performance, the batch normalisation technique[22] is used. For the training and evaluation of the network, each dataset is split into training and test datasets with the ratio of 7 to 3. The loss function is the evidence lower bound[14], and the optimisation algorithm is Adam[23] with a learning rate of 0.0001. See Supplementary Figs. 1a, 2a and Supplementary Tables 1, 2 for more details. We used the WI images as the inputs to the feature extraction network to extract defect position and WI information. The definition of the RMS error in evaluation is $\sqrt{\frac{1}{N}\sum_{n=1}^{N}\left\{\frac{1}{3600}\sum_{i=1}^{60}\sum_{j=1}^{60}\left[X_{i,j}^{(n)} - Y_{i,j}^{(n)}\right]^2\right\}}$, where $N$ is the number of samples in the test dataset, $\mathbf{X}^{(n)}$ is the $n$-th numerical-calculated WI image input, $\mathbf{Y}^{(n)}$ is the $n$-th output image of the feature extraction network [(A) + (B) in Fig. 1d], and the suffixes $i,j$ represent the pixel index. The RMS error is a dimensionless quantity because the WI images $\mathbf{X}^{(n)}$ and $\mathbf{Y}^{(n)}$ are dimensionless quantities that satisfy the normalization conditions: $\sum_{i=1}^{60}\sum_{j=1}^{60} X_{i,j}^{(n)} = 1$ and $\sum_{i=1}^{60}\sum_{j=1}^{60} Y_{i,j}^{(n)} = 1$.

**Geometry generative network and its training.** The geometry generative network is a combination of a fully-connected network and the decoding network in the feature extraction network. The fully-connected network has 4 layers, where the dropout technique[24] is used to improve the learning performance. The loss function is the mean squared error. For the training and evaluation of the network, the dataset is split into training and test datasets (corresponding WIs of the test data are not seen by the QGD network). See

Supplementary Figs. 1b, 2b and Supplementary Table 3 for more details. The definition of the RMS error in the evaluation is the same as that used in the feature extraction network, except that the $\mathbf{Y}^{(n)}$ is replaced by the generated WI image from the *n*-th magneto-conductance input. The definition of the absolute difference image $\mathbf{X}^{\text{dif}}$ of the WI images between $\mathbf{X}$ and $\mathbf{X}'$ is $X_{i,j}^{\text{dif}} = |X_{i,j} - X'_{i,j}|$. Although reconstructing a Hamiltonian from wave functions (an inverse problem) is difficult, from a theoretical point of view, potential reconstruction is possible if scattering states are known for all the incident wavenumbers[25]. In the present case, although the incident wavenumbers are limited to those on the Fermi surface, the spatial distribution of the potential is reproduced. This might be attributed to the fact that the potential shape is restricted to the antidot shape.

**Latent-space dimension in QGD.** To determine the appropriate dimension of the latent space, we performed control experiments. Supplementary Fig. 5 shows the latent-space dimension dependence of the averaged RMS error of the decoded WI images. For the feature extraction network, the latent-space dimension less than 5 was found to show large RMS errors, which can be attributed to insufficient capacity of the latent space as shown in Supplementary Fig. 5a, b. On the other hand, higher latent-space dimension than 9 was found to show large RMS errors due to too many learning parameters. Due to the competition between the two, the RMS error takes a minimum around at dimension = 7. A similar behaviour was also found for the WI images decoded from the geometry generative network (Supplementary Fig. 5c, d). Therefore, we determined to use the 7-dimensional latent space for the QGD network.

**Generalisation ability of the QGD network.** In Supplementary Figs. 6 and 7, we show some training results of the network with different tight-binding parameters: the Fermi energy and disorder potential, respectively. The result shows that the network we proposed exhibits high fidelity regardless of the parameters. Although the interference fringe patterns in the WI images largely depend on the Fermi energy and the disorder potential (see Supplementary Figs. 6 and 7 and those caption for details), the method can reconstruct WI images. In Supplementary Fig. 8, we also show the WI images decoded from the Fermi energy dependence of the magneto-conductance, where the Fermi energy is swept from 1.2*t* to 2.0*t* at zero magnetic flux density. The data shows that the WI images can also be decoded with high fidelity when we use the Fermi energy dependence data instead of the magnetic flux density dependence.

**Quantitative analysis of the data structure in the latent space.** We quantitatively estimated the difference between Fig. 3c and d by calculating the thicknesses of the data structures as the variance of the data points in the thickness direction. Firstly, as shown in Supplementary Fig. 11a, we cut out a part of the data structure in the latent space, which locally forms two-dimensional plane with a thickness. The local variance was calculated for the in-plane and thickness directions. The ratio of the local variance along the in-plane and thickness directions is 0.10 for the data structure formed by the geometry images with WIs (Supplementary Fig. 11c). On the other hand, the data structure formed by the geometry images without WIs exhibits a much smaller value of the ratio, 0.01 (Supplementary Fig. 11b and d).

**Dual-layered data structure in the latent space.** In the present calculation, we chose the pixel size of the interference images so that the Fermi wavelength is equivalent to 2 pixels. In the condition, the data scattering in the latent space was found to apparently be quantised to show the most interpretable structure: the dual layers, which can be explained by the fact that a standing wave between an antidot and the sample end can be classified into two; destructive and constructive interference patterns. In fact, the dual-layer structure was found to change into a randomly scattered structure around the surface by changing the pixel size, showing that the essential feature here is the data scattering around the surface structure representing the information of the antidot positions.

**Quantitative comparison between original and deciphered WI images.** We analysed the WI images in terms of the normalised cross correlation:

$$R_{\text{NCC}} = \frac{\sum_{i,j} A(i,j) B(i,j)}{\sqrt{\sum_{i,j} A(i,j)^2 \sum_{i,j} B(i,j)^2}},$$

where **A** and **B** are the original and generated WI images, respectively. As shown in Supplementary Fig. 9a, b, $R_{\text{NCC}}$ increases with the training epoch. $R_{\text{NCC}}$ after the training is much greater ($R_{\text{NCC}} > 0.997$) than that before the training. We confirmed that the generated WI images show high fidelity in terms of the normalised cross correlation.

**Data availability.** The data that support the findings of this study are available from the corresponding author upon reasonable request.

**Code availability.** The codes that support the findings of this study are available from the corresponding author upon reasonable request.

**Acknowledgments** This is a post-peer-review, pre-copyedit version of an article published in Nature Communications. The final authenticated version is available online at: http://dx.doi.org/10.1038/s41467-022-30767-w. The authors thank Y. Oikawa for valuable discussions. This work was supported by ERATO "Spin Quantum Rectification Project" (No. JPMJER1402) from JST, Japan; Grant-in-Aid for Scientific Research (S) (No. JP19H05600), (A) (No. JP19H00658), (B) (No. JP20H02599), and (C) (No. JP20K05297), Grant-in-Aid for Early-Career Scientists (No. 21K14519), Grant-in-Aid for Research Activity Start-up (Nos.

JP19K21031 and JP19K21035, JP20K22476) from JSPS KAKENHI, Japan; Institute for AI and Beyond of the University of Tokyo, Japan; and the NEC Corporation. R.R. acknowledges support from the European Commission through the project 734187-SPICOLOST (H2020-MSCA-RISE-2016), the European Union's Horizon 2020 research and innovation program through the MSCA grant agreement SPEC-894006, Grant RYC 2019-026915-I funded by the MCIN/AEI/10.13039/501100011033 and by "ESF investing in your future", the Xunta de Galicia (ED431B 2021/013, Centro Singular de Investigación de Galicia Accreditation 2019-2022, ED431G 2019/03) and the European Union (European Regional Development Fund - ERDF). K.O. acknowledges support from GP-Spin at Tohoku University.

**Author contributions** S.D. and K.T. contributed equally to this work. E.S. planned and supervised the study. S.D. and K.T. performed numerical calculations, analysed the data, and prepared the manuscript with help from T.O. S.K. built and optimised the machine learning models. T.K., R.R., K.O., T.O. and E.S. discussed the results, developed the explanation of the experiments, and commented on the manuscript.

**Competing interests** The authors declare no competing interests.

**Additional information**

Correspondence and requests for materials should be addressed to S.D.

Reprints and permissions information is available at http://www.nature.com/reprints.

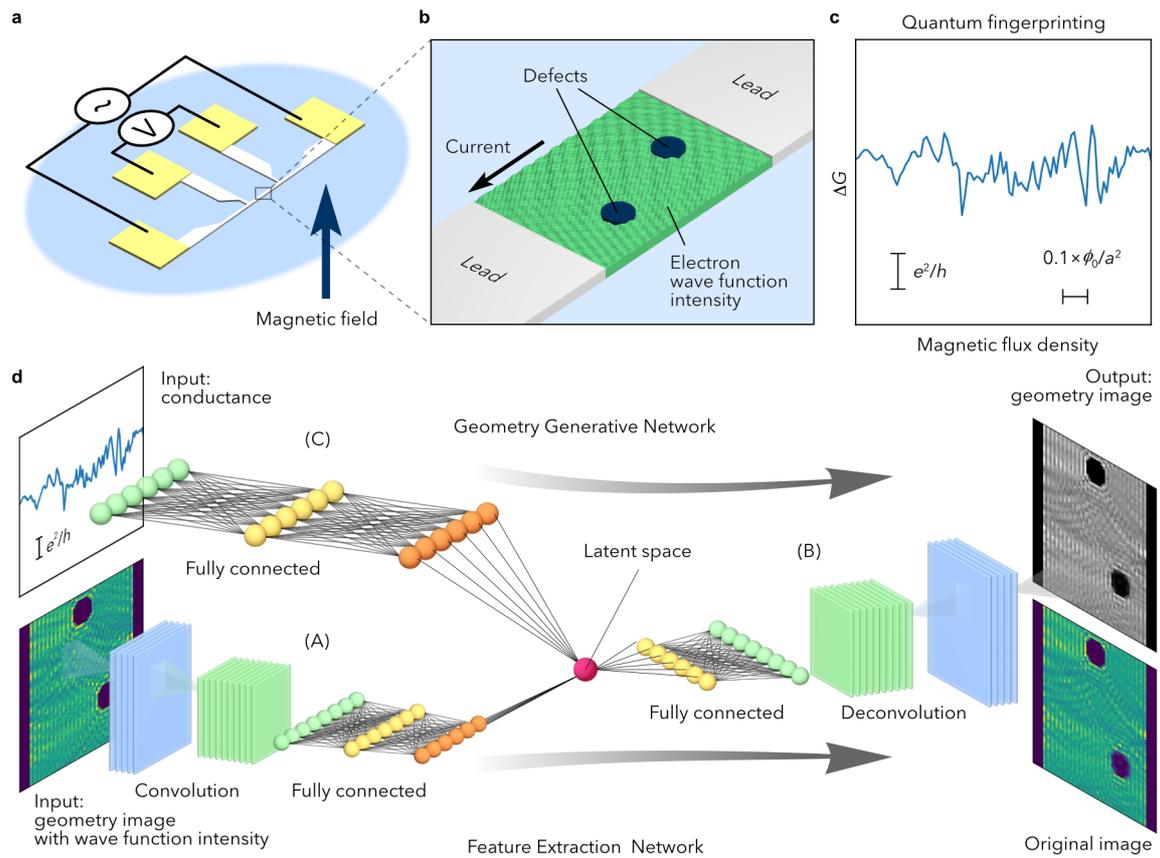

**Fig. 1 Concept of Quantum Geometric Decoder. a, b** A schematic illustration of a magneto-conductance measurement in a small metal sample (**a**) and its magnified view (**b**). The green stripe pattern describes electron wave function intensity (WI) in the sample with defects. **c** Conductance change $\Delta G$ for a nanowire sample. $e$, $h$, and $a$ are the elementary charge, the Plank constant, and the lattice constant in the calculation model, respectively. $\phi_0 = h/e$ is the magnetic flux quantum. **d** Concept of Quantum Geometric Decoding based on a deep neural network. First, the networks (A) and (B) compress the WI images into the latent space. Then, the networks (C) and (B) output a geometry image including the sample shape, defect distribution, and WI information from the input of the magneto-conductance.

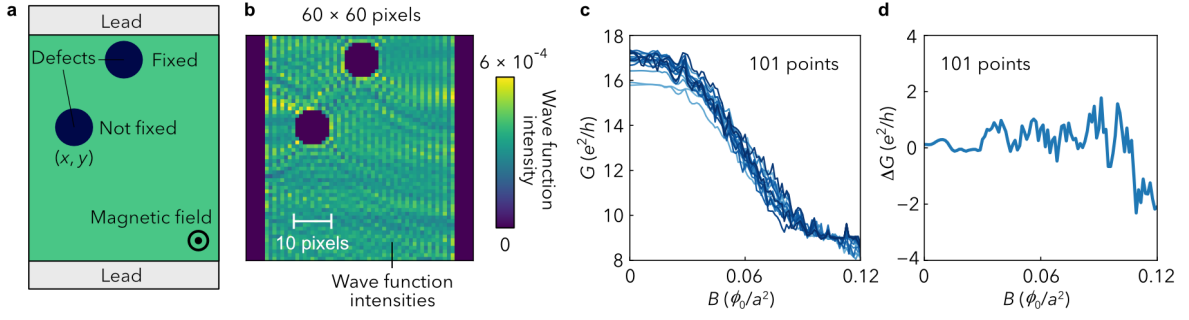

**Fig. 2 Numerical calculation of WI distribution and magneto-conductance in a sample with antidot defects. a** A sample system for calculation with two antidot defects. One antidot is fixed at the upper centre of the nanowire, while the other is located so as not to overlap with the fixed one. Two leads are attached to the top and bottom ends of the sample to measure conductance. **b** Calculated WI distribution. The square of the absolute value of the calculated wave function is plotted in the sample region with 60 × 50 pixels. The WI values are normalised such that the sum of the intensity equals to unity for each antidot configuration; $\sum_{i=1}^{60}\sum_{j=1}^{60} X_{i,j} = 1$, where **X** is a WI image, and the suffixes $i, j$ represent the pixel label. We added zero padding with 5 sites to the left and right ends of the nanowire. **c** Magneto-conductance $G$ for 10 samples with different defect distributions. Here, $B$ is the magnetic flux density. **d** Normalised data of the calculated magneto-conductance $\Delta G$ for the antidots distribution shown in **a**. $\Delta G$ is obtained by subtracting the averaged $G$ ($\equiv G_{\text{ave}}$) over all the nanowire configurations from the $B$ dependence of $G$ [$\Delta G(B) \equiv G(B) - G_{\text{ave}}(B)$].

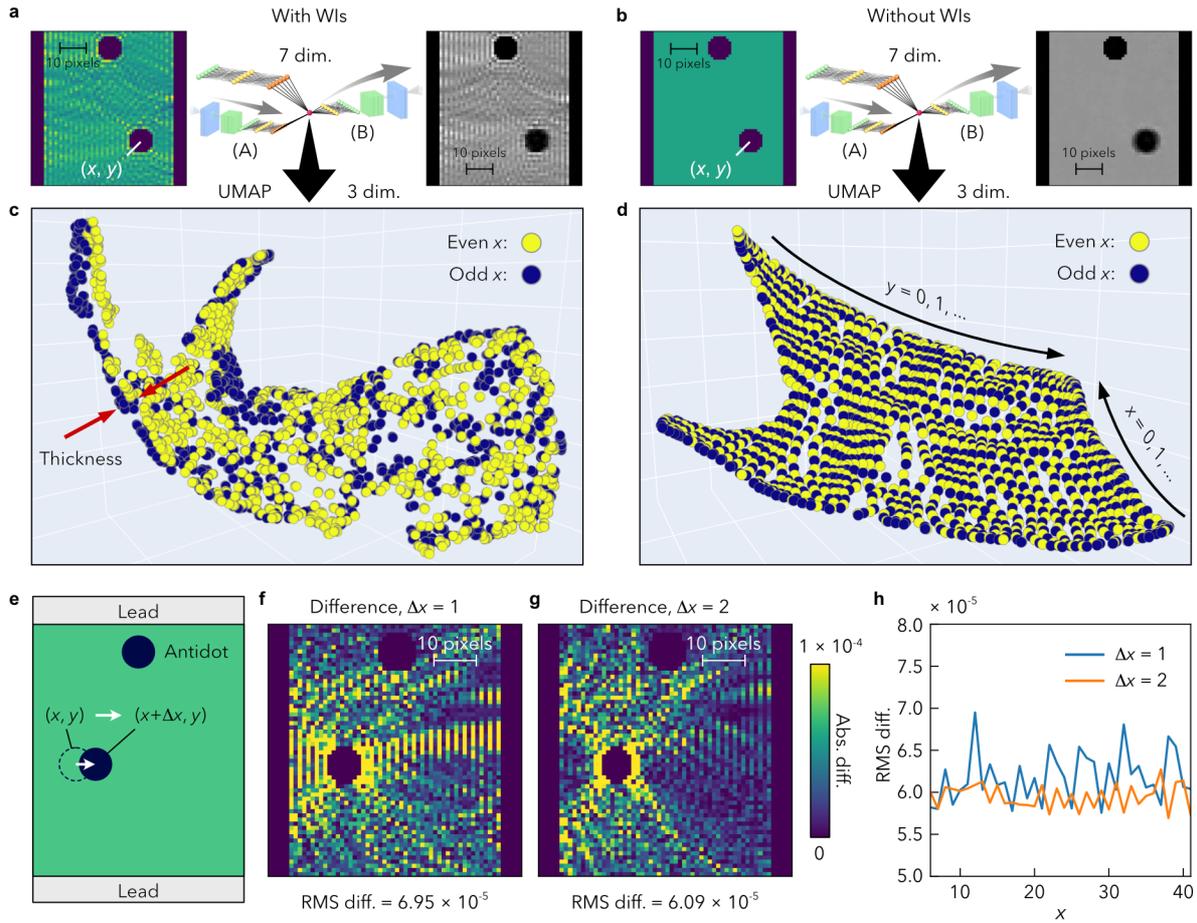

**Fig. 3 Visualisation of the data geometric structure in the feature extraction network. a, b** The feature extraction network based on VAE, trained by using geometry images with WIs (**a**) and without WIs (**b**), and an example of input and output images. **c, d** UMAP data points generated in the latent space, trained with the geometry images together with WIs (**c**), and without WIs (**d**). $(x, y)$ represents the location of the antidot defect. The data structure is mapped onto a 3-dimensional space from the 7-dimensional latent space by using UMAP to visualise the data geometry. Each data point corresponds to one WI image and is coloured in blue or yellow depending on the parity of $x$. **e** The definition of $\Delta x$. **f, g** Images of the absolute difference between $(x, y)$ data and $(x + \Delta x, y)$ data, where $(x, y) = (12, 33)$, for $\Delta x = 1$ (**f**) and $\Delta x = 2$ (**g**), respectively. The RMS difference values are shown below the images. **h** The RMS differences at $(x, 33)$.

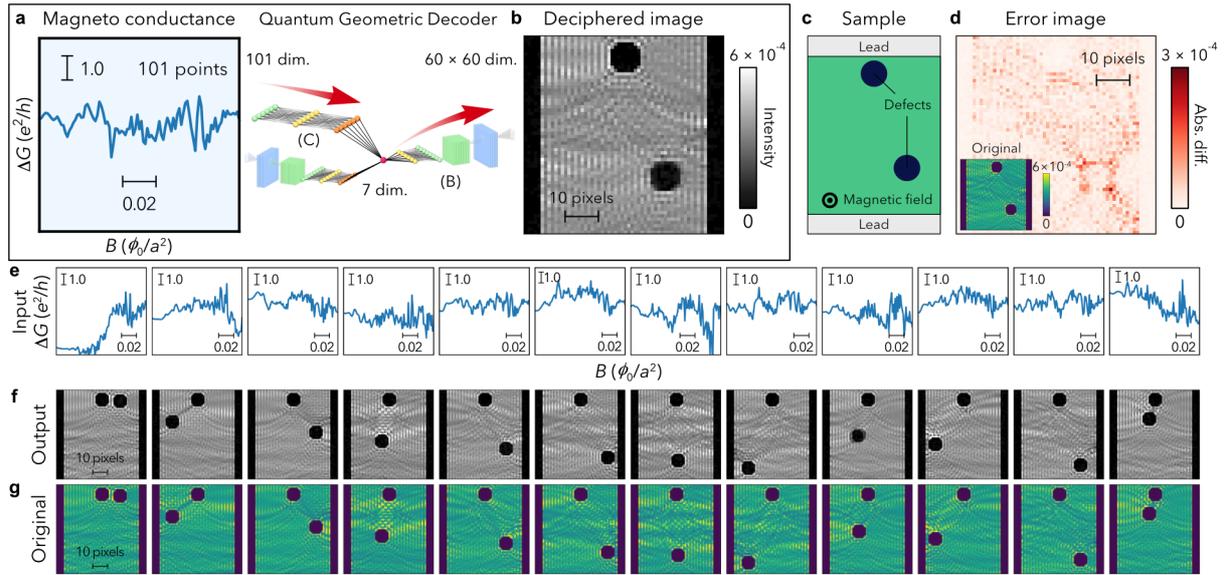

**Fig. 4 Result of Quantum Geometric Decoding (QGD). a, b** The geometry generative network and an example of the input conductance and output deciphered image. $\Delta G$ and $B$ are the normalised magneto-conductance and the magnetic flux density, respectively. **c** The sample system whose magneto-conductance is shown in **a**. **d** The absolute difference image between the original image (inset) and the output deciphered image shown in **b**. **e, f, g** Examples of the input conductance (**e**), output deciphered images (**f**), and original images (**g**). Colour scales of the intensity images in **f** and **g** are the same as those of **b** and the inset to **d**, respectively.

# Supplementary Information

# Deciphering quantum fingerprints in electric conductance

S. Daimon, K.  Tsunekawa, S. Kawakami, T. Kikkawa, R. Ramos, K. Oyanagi, T. Ohtsuki & E. Saitoh

## Table of Contents

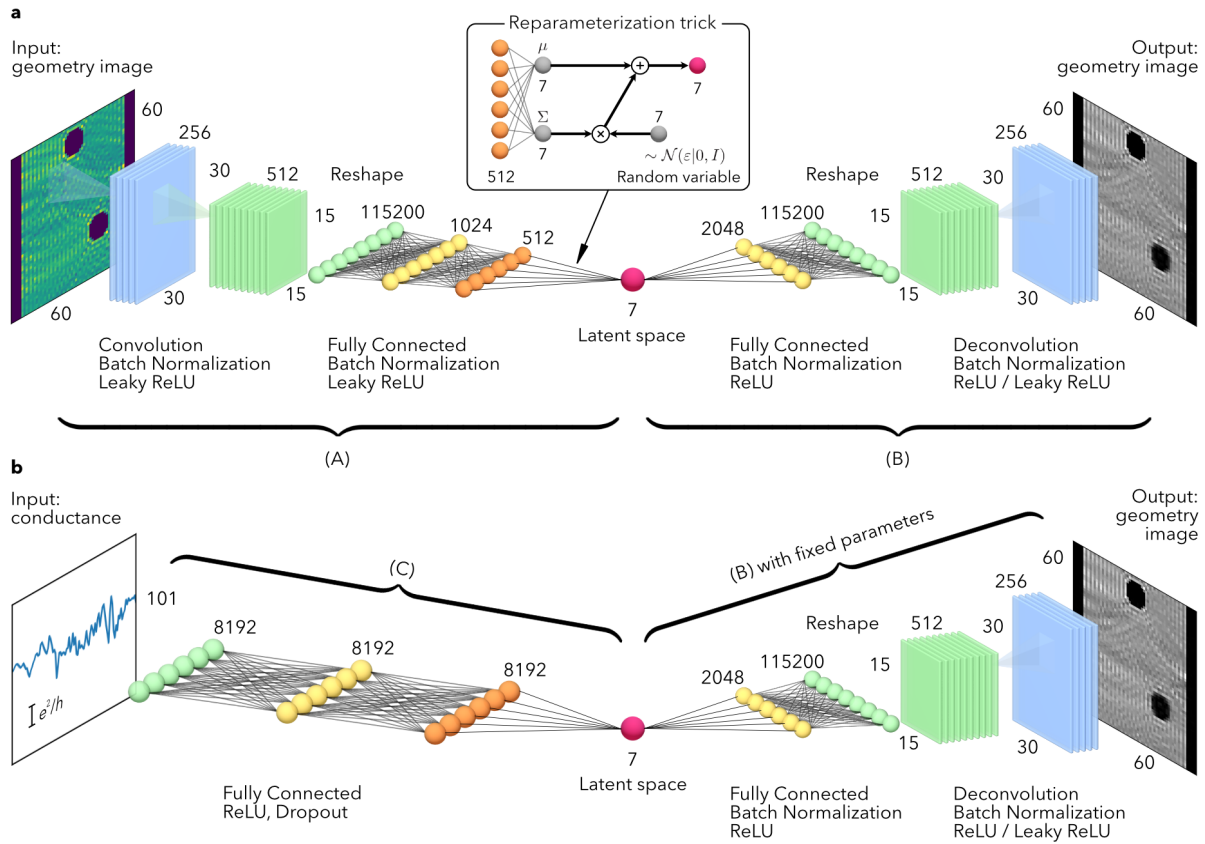

**Supplementary Figure 1 | Architecture of the feature extraction and geometry generative networks. a** The architecture of the feature extraction network based on the variational autoencoder[1]. The feature extraction network consists of a convolution part and a deconvolution part, labeled as (A) and (B), respectively. The convolution part is composed of 2 two-dimensional convolution layers and 3 fully-connected layers (see Supplementary Table 1 for more details). The input mono-channel image size is 60 × 60, and the output is the mean value **μ** and the logarithm variance **Σ** in the 7-dimensional latent space. A 7-dimensional vector in the latent space is randomly selected in the reparameterization trick[1,2]. The deconvolution part is composed of 2 fully-connected layers and 2 two-dimensional transposed convolution layers (see Supplementary Table 2 for more details). The input is a 7-dimensional vector in the latent space, and the output mono-channel image size is 60 × 60. We used rectified linear units (ReLUs)[3] or leaky ReLUs[4] for activation functions and the batch normalisation technique[5] in the feature extraction network. The input wave function images are multiplied by 5 as a preprocess for a training. **b** The architecture of the geometry generative network. The geometry generative network consists of a prediction part and the deconvolution part, labeled as (C) and (B), respectively. The prediction part is composed of 4 fully-connected layers with ReLU activation functions and the dropout technique[6] (see Supplementary Table 3 for more details). The

input is normalised magneto-conductance data with 101 points, and the output is a 7-dimensional vector in the latent space. The deconvolution part is the same as that of the feature extraction network.

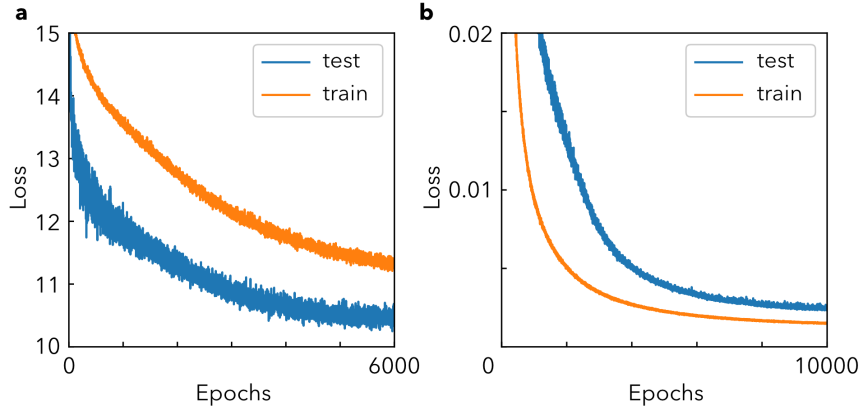

**Supplementary Figure 2 | Learning curves. a** Loss values for the training and test datasets at each epoch are plotted for the feature extraction network. The network is trained by the adaptive moment estimation[7] (Adam) with the beta hyperparameters of (0.5, 0.999) and the learning rate of 0.0001. The loss function is the evidence lower bound[1] $L = \frac{1}{N}\sum_{n=1}^{N}\{SE_n + D_{KL}[N(\mathbf{z}; \boldsymbol{\mu}_n, \boldsymbol{\Sigma}_n) || N(\mathbf{z}; \mathbf{0}, \mathbf{I})]\}$, where $n$ is the sample labeling number, $N$ is the number of the samples in the dataset, $SE_n$ is the squared error between the $n$-th input image $\mathbf{X}^{(n)}$ and the $n$-th output image $\mathbf{Y}^{(n)}$; $SE_n = \sum_{i=1}^{60}\sum_{j=1}^{60}\left[X_{i,j}^{(n)} - Y_{i,j}^{(n)}\right]^2$, the suffixes $i, j$ representing the pixel, $D_{KL}$ is the Kullback-Leibler divergence, $N(\mathbf{z}; \boldsymbol{\mu}, \boldsymbol{\Sigma})$ is a Gaussian distribution with its mean value $\boldsymbol{\mu}$ and covariance matrix $\boldsymbol{\Sigma}$, and $\mathbf{I}$ is the identity matrix. Batch size for the training is 256. The test loss is less than the training loss since the reparameterization trick is not used for the test loss evaluation. **b** Loss values for the training and test datasets at each epoch are plotted for the geometry generative network. The network is trained by the adaptive moment estimation[7] (Adam) with the beta hyperparameters of (0.5, 0.999) and the learning rate of 0.00001. The loss function $L$ is the mean squared error of the points in the latent space; $L = \frac{1}{N}\sum_{n=1}^{N}\left\|\mathbf{z}^{(n)} - \mathbf{y}^{(n)}\right\|^2$, where $N$ is the number of samples in the dataset, $\mathbf{y}^{(n)}$ is the output of the network (C) [Supplementary Fig. 1b] from the $n$-th magneto-conductance input, $\mathbf{z}^{(n)}$ is the outputs of the trained network (A) [Supplementary Fig. 1a] from the $n$-th wave function image input. The batch size for the training is 256. The dropout technique is not used for the test loss evaluation.

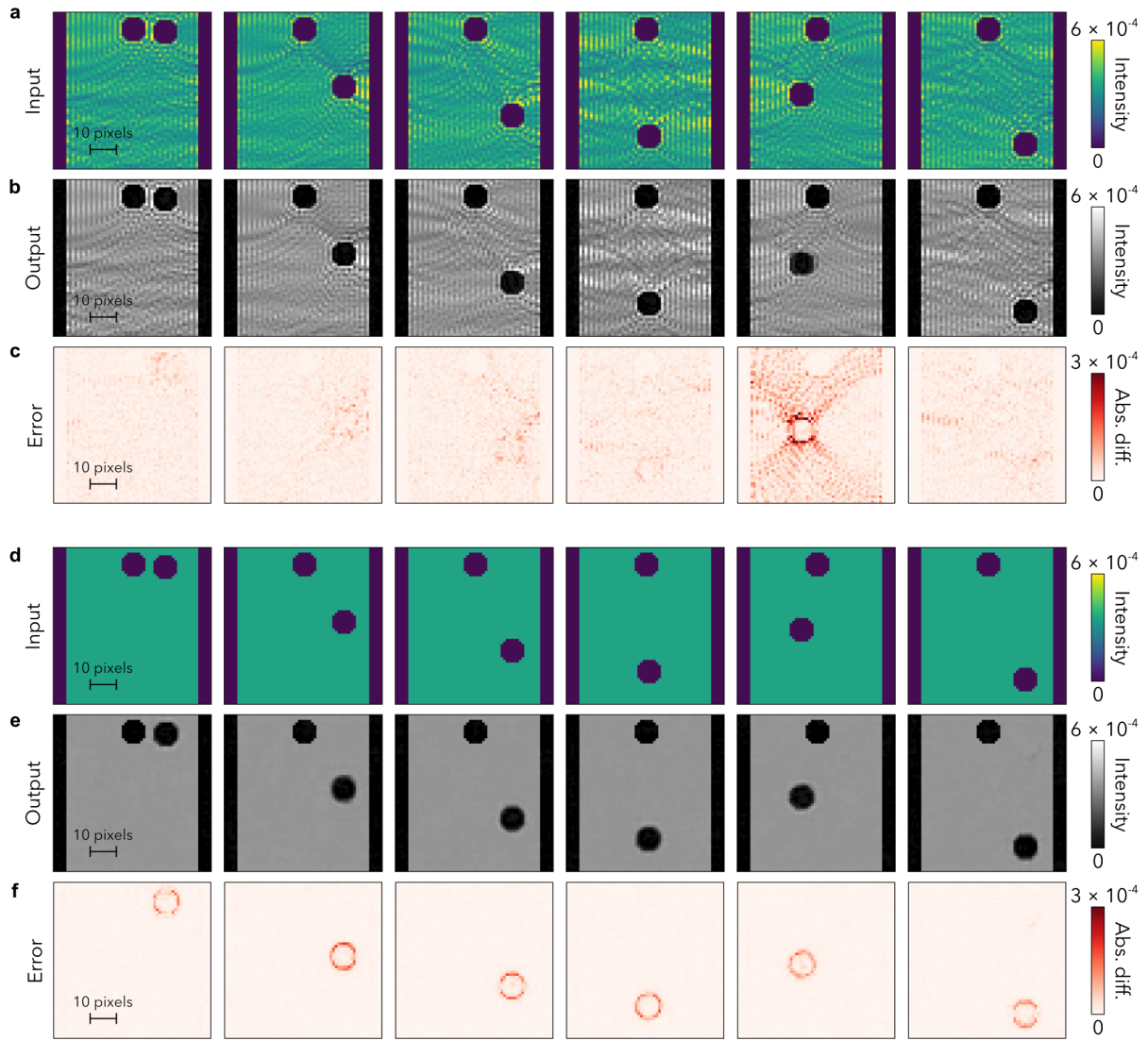

**Supplementary Figure 3 | Autoencoding of geometry images using the feature extraction network after training.** Examples of the input, output, and error images for the feature extraction network (Supplementary Fig. 1a), trained by the geometry images with wave function intensities (WIs) (**a**, **b**, **c**) and without WIs (**d**, **e**, **f**). The input images are randomly selected from the test dataset, not used for the training of the network. The error is defined as the absolute difference between the values of each pixel in the original and output images.

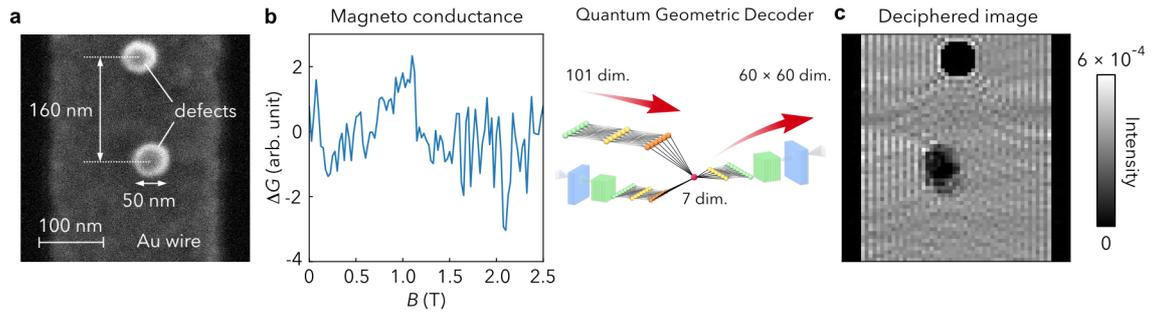

**Supplementary Figure 4 | An experiment using a real nano system. a** A scanning electron microscope image of an Au wire fabricated by electron beam lithography. Black and gray areas are the $SiO_2$ substrate and Au wire, respectively. The white circles in the wire are the damaged areas introduced by electron beam exposure. **b** The magnetic field $B$ dependence of the normalised magneto-conductance $\Delta G$ measured by a four probe measurement at 100 mK (see Supplementary Note 1 for details). **c** The output of the QGD network for the data in **b**.

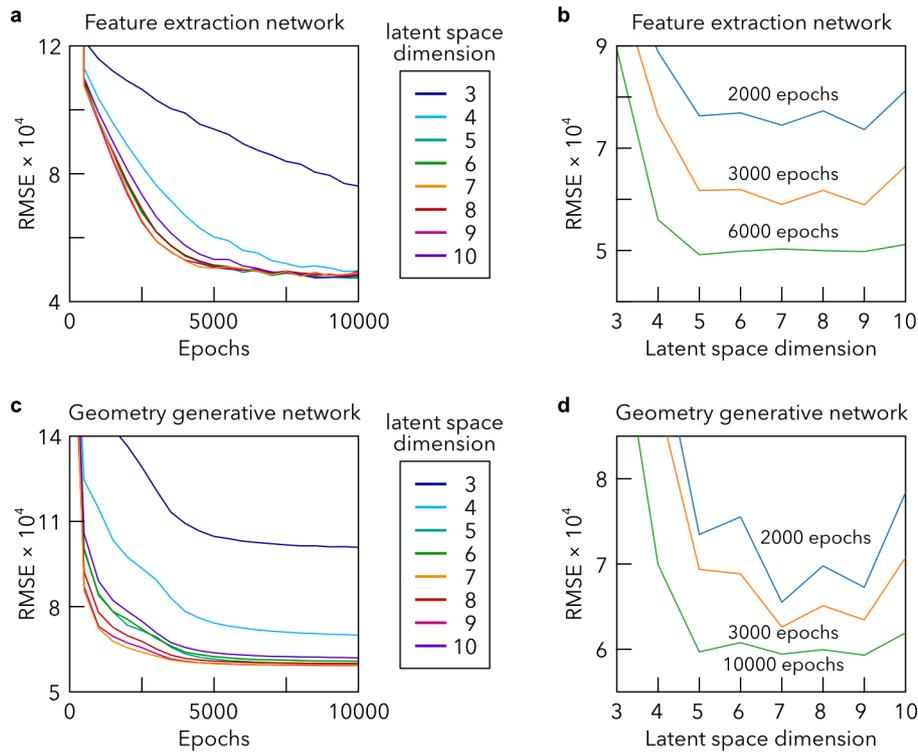

**Supplementary Figure 5 | The latent-space dimension dependence of the training results. a** The training epoch dependence of the RMS error between the original and deciphered images generated from the feature extraction network. The training results for different latent-space dimensions are plotted by different colours. **b** The latent-space dimension dependence of the RMS error between the original and deciphered images generated from the feature extraction network. The training results for 2000, 3000, and 6000 epochs are plotted by blue, orange, and green lines. **c** The training epoch dependence of the RMS error between the original and deciphered images generated via the geometry generative network. The training results for different latent-space dimensions are plotted by different colours. **d** The latent-space dimension dependence of the RMS error between the original and deciphered images generated via the geometry generative network. The training results for 2000, 3000, and 10000 epochs are plotted by blue, orange, and green lines.

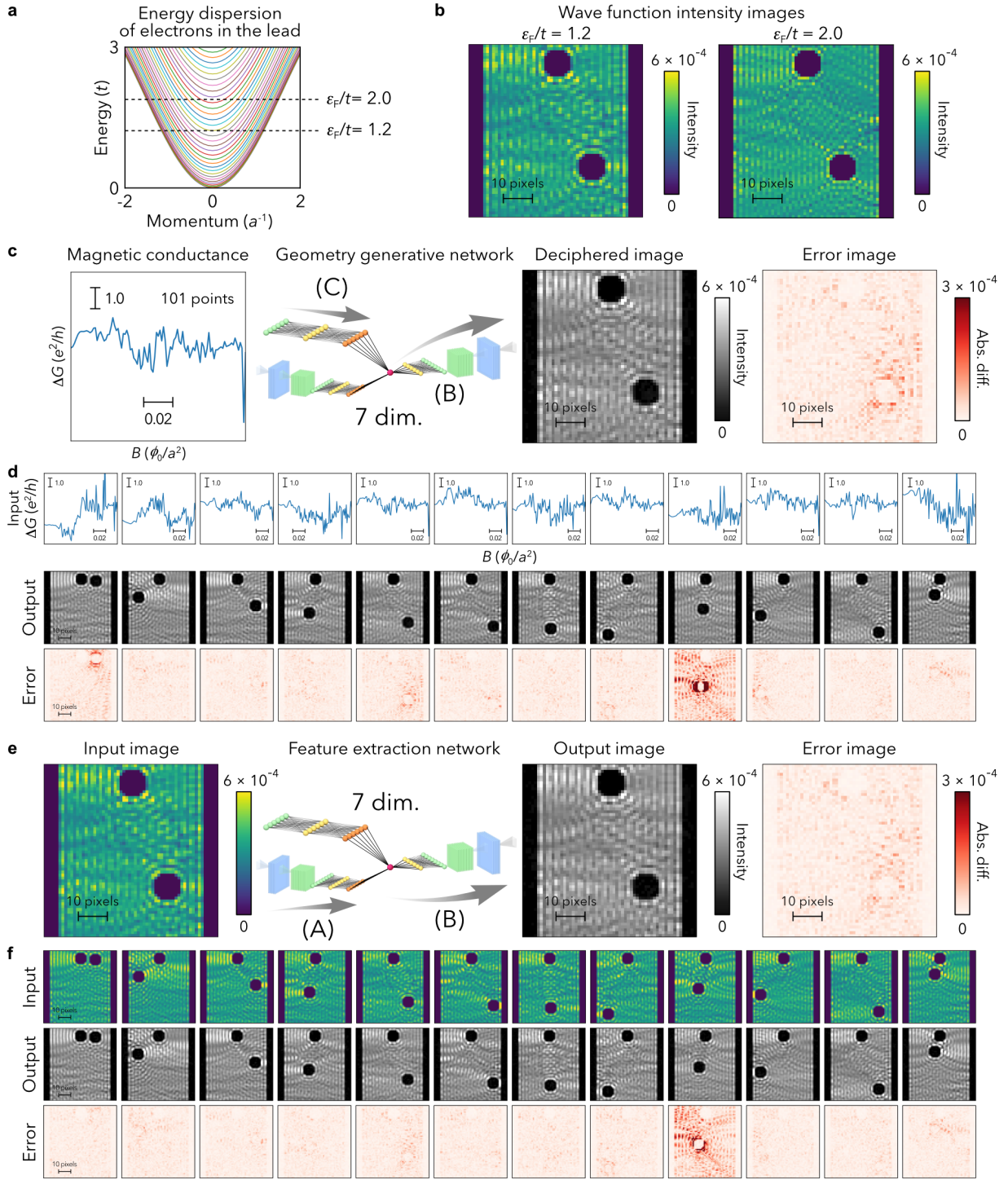

**Supplementary Figure 6 | Result of Quantum Geometric Decoding using data with different Fermi energy. a** The energy dispersion relation of electrons in the lead. $t$, $\varepsilon_F$, and $a$ are the hopping energy, Fermi energy, and lattice constant, respectively. **b** Wave function intensity images for $\varepsilon_F/t = 1.2$ and 2.0. We used $\varepsilon_F/t = 1.2$ for the machine learning shown in **c**-**f** while we used $\varepsilon_F/t = 2.0$ for the experiments in the main text. **c** The geometry generative network and an example of the input conductance, output deciphered image, and absolute difference between the original and deciphered images. $\Delta G$ and $B$ are the normalised magneto-

conductance and the magnetic flux density, respectively. $\phi_0 = h/e$ is the magnetic flux quantum. **d** Examples of the input conductance, output deciphered images, and error images. **e** The feature extraction network and an example of the input WI image, output deciphered image, and absolute difference between the original and deciphered images. **f** Examples of the input images, output deciphered images, and error images.

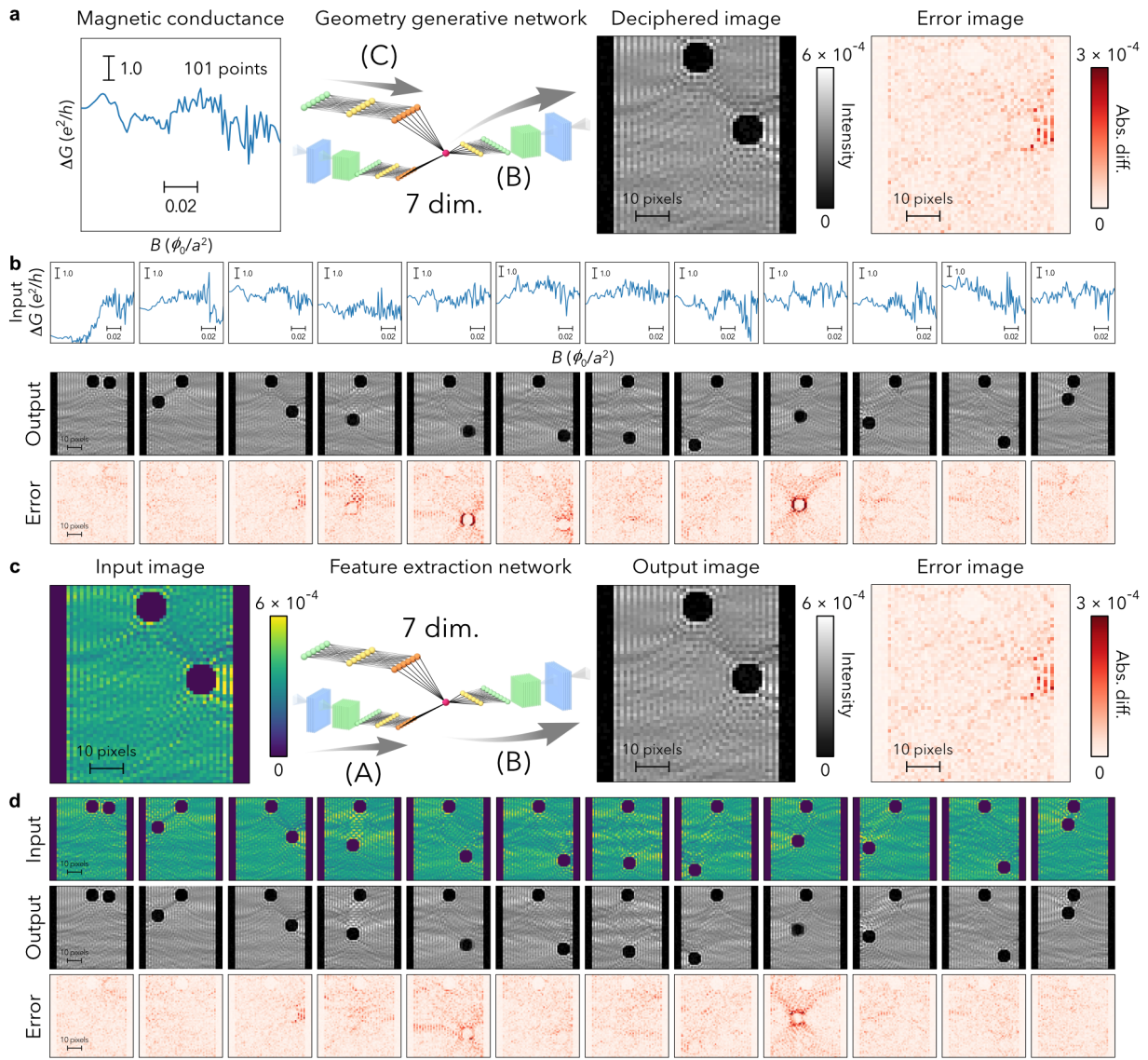

**Supplementary Figure 7 | Result of Quantum Geometric Decoding using data with greater disorder potentials. a** The geometry generative network and an example of the input conductance, output deciphered image, and absolute difference between the original and deciphered images. Here, the disorder-potential amplitude is doubled compared to that used for the dataset preparation in the main text. **b** Examples of the input conductance, output deciphered images, and error images. **c** The feature extraction network and an example of the input WI image, output deciphered image, and absolute difference between the original and deciphered images. **d** Examples of the input images, output deciphered images, and error images.

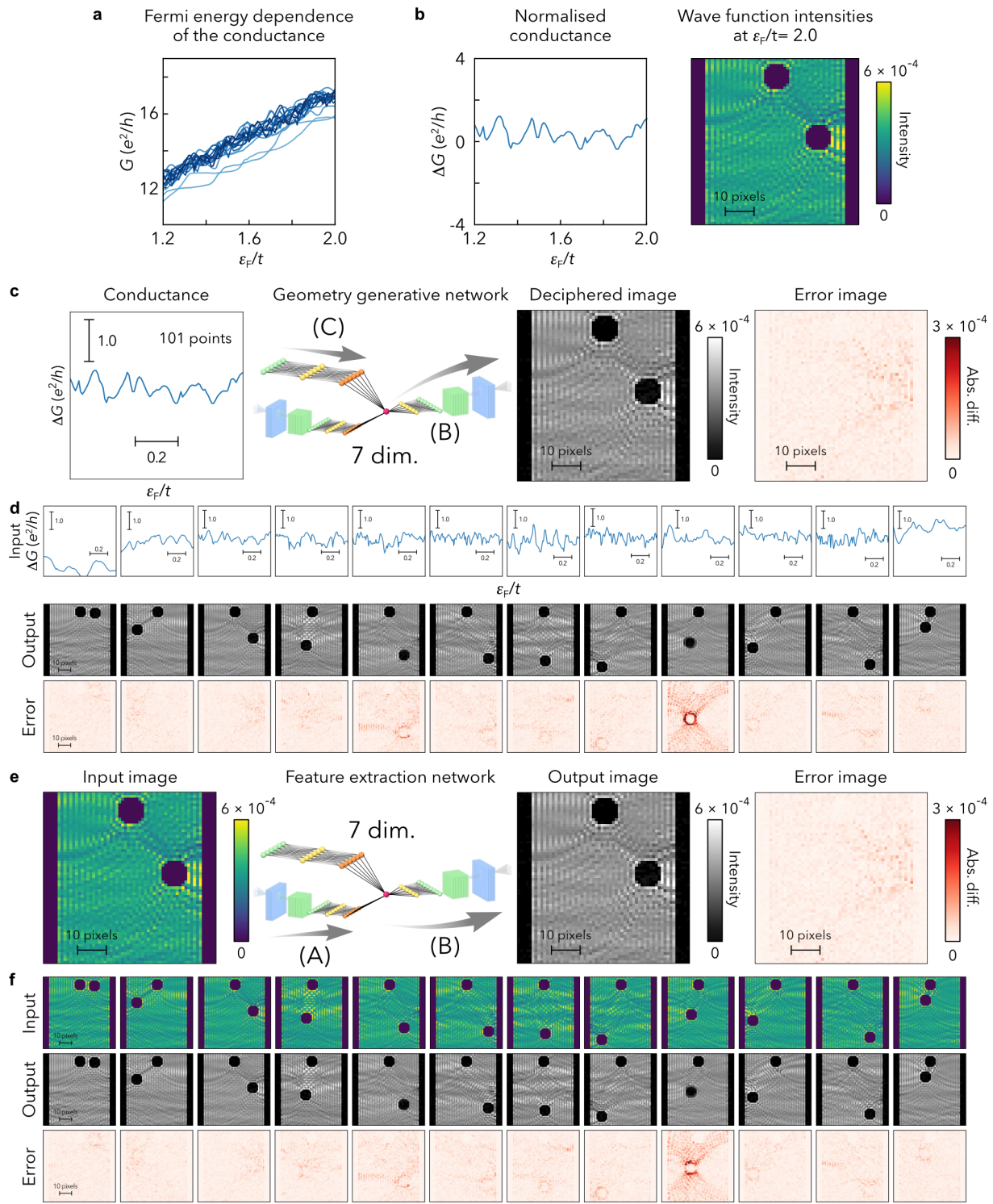

**Supplementary Figure 8 | Result of Quantum Geometric Decoding using the Fermi-energy dependence of conductance.** **a** The Fermi energy dependence of the conductance at zero magnetic field and $\varepsilon_F/t = 2.0$. **b** Normalised conductance and a WI image used for the training. **c** The geometry generative network and an example of the input conductance, output deciphered image, and absolute difference between the original and deciphered images. **d** Examples of the input conductance, output deciphered images, and error images.

**e** The feature extraction network and an example of the input WI image, output deciphered image, and absolute difference between the original and deciphered images. **f** Examples of the input images, output deciphered images, and error images.

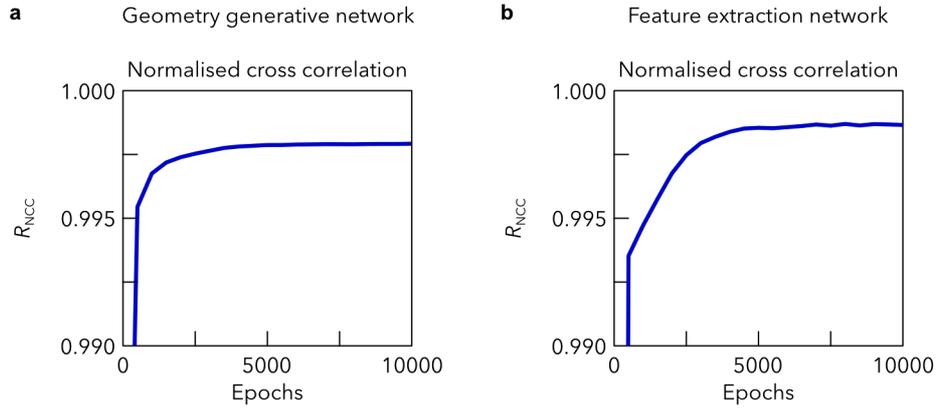

**Supplementary Figure 9 | Quantitative analysis of the deciphered WI images. a, b** The training epoch dependence of the normalised cross correlation:

$$R_{\text{NCC}} = \frac{\sum_{i,j} A(i,j)B(i,j)}{\sqrt{\sum_{i,j} A(i,j)^2 \sum_{i,j} B(i,j)^2}}$$

between the original WI image (**A**) and the deciphered WI image (**B**) for the geometry generative network (**a**) and the feature extraction network (**b**). We confirmed that the generated WI images show high fidelity ($R_{\text{NCC}} > 0.997$) in terms of the normalised cross correlation.

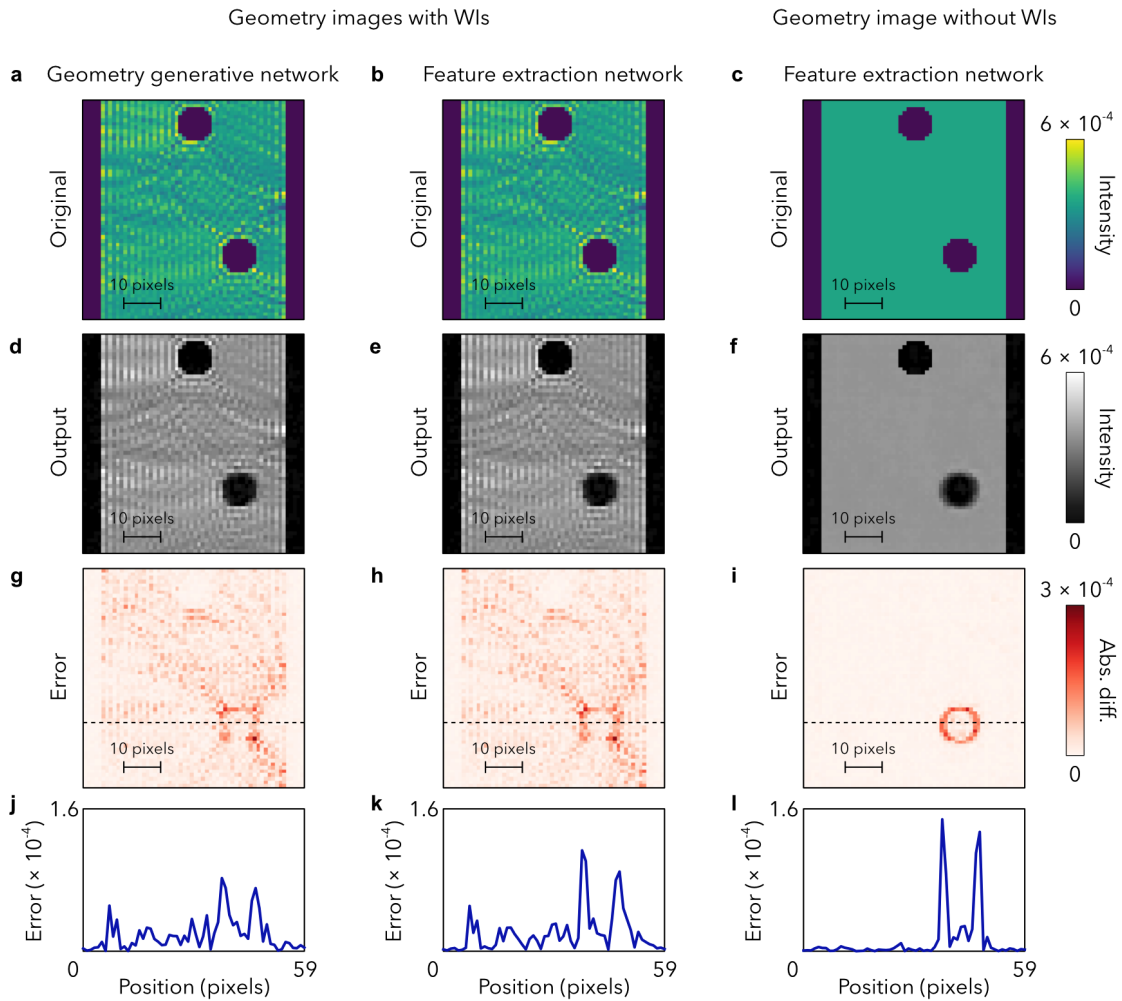

**Supplementary Figure 10 | The position dependence of the absolute difference between the original and deciphered geometry images. a**, **b**, **c** The input geometry images with WIs for the geometry generative network (**a**) and the feature extraction network (**b**), and the input geometry image without a WI for the feature extraction network (**c**). **d**, **e**, **f** The corresponding output geometry images. **g**, **h**, **i** The absolute difference (error) images between the input and output images corresponding to the data in **a**, **b**, and **c**. **j**, **k**, **l** The one-dimensional position dependence of the absolute difference (error) between the input and output images. In the panels **j**, **k**, and **l**, we plotted the error values along the dotted lines in **g**, **h**, and **i**, respectively.

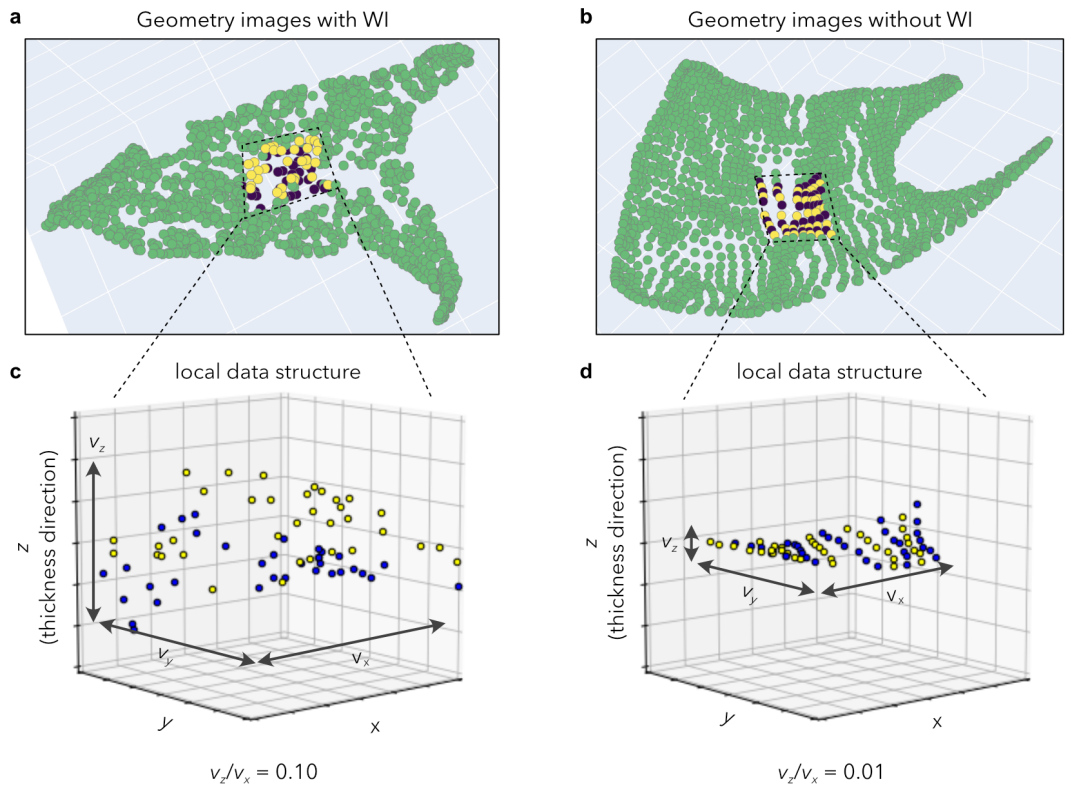

**Supplementary Figure 11 | Dispersion analysis on the data generated from geometry images with and without the WI information. a, b** Data structures generated in the latent space trained with the geometry images together with WIs (**a**), and without WIs (**b**). **c, d** Local structures cut out from the data shown in **a** and **b**, respectively. The $x$, $y$, and $z$ axes are chosen so as to maximise $v_x$ and then maximise $v_y$, where $v_x$, $v_y$, and $v_z$ are the local variance of the data structure along the $x$, $y$, and $z$ axes, respectively. $v_z/v_x$ are 0.10 and 0.01 for the data structure in **c** and **d**, respectively.

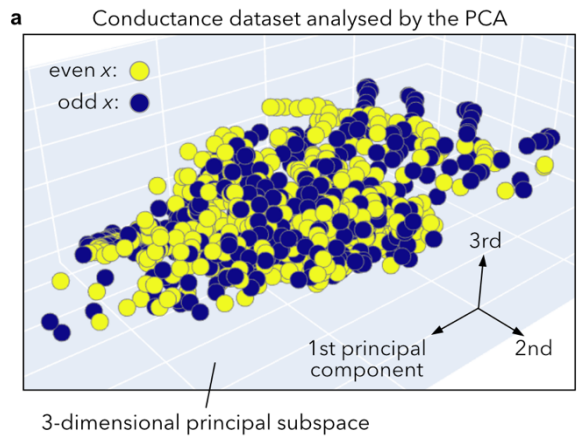 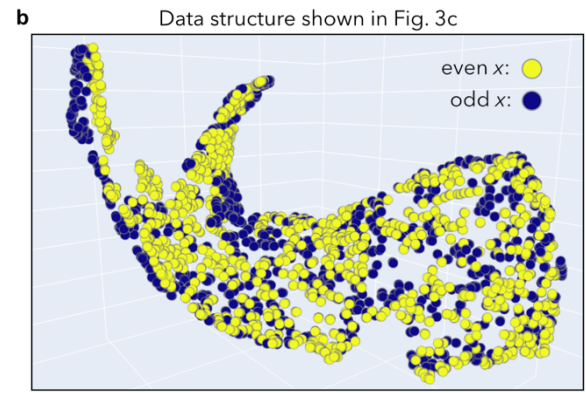

**Supplementary Figure 12 | Principal component analysis (PCA) on the conductance traces. a** Output from the PCA analysis on the conductance plotted in the 3-dimensional principal subspace. **b** Data structure shown in Fig. 3c (output of the feature extraction network).

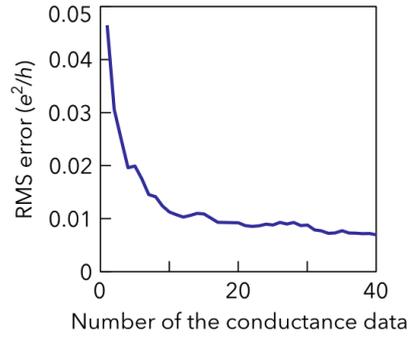

**Supplementary Figure 13 | Convergence of the conductance data.** We calculated 500 conductance traces with different disorder configurations with fixed antidot positions and obtained the averaged conductance $G_{AVE}$. To check the convergence of the averaged conductance, we plotted the RMS error: $\sqrt{\sum_{i=1}^{101}|G_{n,i} - G_{AVE,i}|^2 / 101}$, where $n$ is the number of the disorder configurations and $G_n$ is the disorder averaged conductance averaged over $n$ conductance traces. $i$ is the label representing the magnetic field. The RMS error rapidly decreases with increasing the number of disorder configurations.

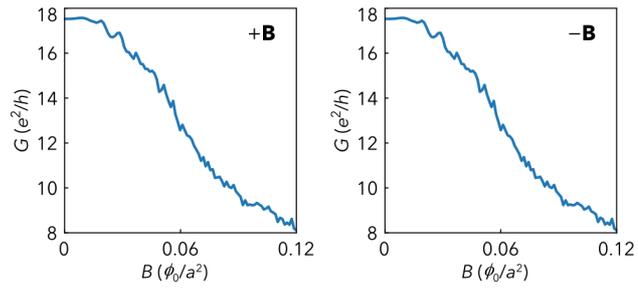

**Supplementary Figure 14 | Comparison between the calculated magnetic fingerprints with opposite magnetic field directions.** We checked whether the relationship required by the symmetry is satisfied. The same magnetic fingerprint appears with respect to the magnetic field reversal.

**Supplementary Table 1 | Architecture of the convolution part of the feature extraction network.**

| Layer # | Layer name | Hyperparameters |
|---|---|---|
| 1 | Convolution | output channel=256, kernel size=4, stride=2, padding=1 |
| 2 | Leaky ReLU | negative slope=0.2 |
| 3 | Convolution | output channel=512, kernel size=4, stride=2, padding=1 |
| 4 | Batch normalisation | |
| 5 | Leaky ReLU | negative slope=0.2 |
| 6 | Linear | output features=1024 |
| 7 | Batch normalisation | |
| 8 | Leaky ReLU | negative slope=0.2 |
| 9 | Linear | output features=512 |
| 10-1 | Linear (mean value) | output features=7 |
| 10-2 | Linear (logarithm variance) | output features=7 |

**Supplementary Table 2 | Architecture of the deconvolution part of the feature extraction network.**

| Layer # | Layer name | Hyperparameters |
| --- | --- | --- |
| 11 | Linear | output features=2048 |
| 12 | Batch normalisation | |
| 13 | ReLU | |
| 14 | Linear | output features=115200 |
| 15 | Batch normalisation | |
| 16 | ReLU | |
| 17 | Transposed convolution | output channel=64, kernel size=4, stride=2, padding=1 |
| 18 | Batch normalisation | |
| 19 | ReLU | |
| 20 | Transposed convolution | output channel=1, kernel size=4, stride=2, padding=1 |

**Supplementary Table 3 | Architecture of the prediction part of the geometry generative network.**

| Layer # | Layer name | Hyperparameters |
|---------|------------|-----------------|
| 21 | Linear | output features=8192 |
| 22 | ReLU | |
| 23 | Dropout | Ratio=0.5 |
| 24 | Linear | output features=8192 |
| 25 | ReLU | |
| 26 | Dropout | Ratio=0.5 |
| 27 | Linear | output features=8192 |
| 28 | ReLU | |
| 29 | Dropout | Ratio=0.5 |
| 30 | Linear | output features=7 |

**Supplementary Note 1 | Experimental.**

To apply QGD to data from real objects, fine tuning the network by using data from real objects is important, although it requires enormous experimental resources and a long time. In this section, to show the current state of QGD application, we simply show some experimental data.

On a trial basis, we apply the trained QGD without fine tuning to experimental data from a nano-structured Au film. In Supplementary Fig. 4a, we show a scanning electron microscope image for the sample. The sample was fabricated by the lift-off technique using electron beam lithography. A 25 nm-thick Au film with a width of 280 nm is deposited on a $SiO_2$/Si substrate in ultra-high vacuum. The Au wire has two damaged areas with diameters of ~50 nm. The damaged parts are made by irradiating an electron beam with an acceleration voltage of 50 kV and a beam current of 130 pA for 20 seconds. The magneto-resistance was measured by a four probe method with an a.c. resistance bridge (Lake Shore 372). The amplitude and frequency of the applied a.c. charge current were set to 3.16 µA and 11.6 Hz, respectively. The measurement was performed at 100 mK using a $^3$He-$^4$He dilution refrigerator (KelvinoxMX200). The magnetic field was swept with a sweep rate of 12 mT/min. The field pattern of the resistance is almost same between data measured with decreasing and increasing the magnetic field applied perpendicular to the film.

The normalised magneto-conductance $\Delta G$ shown in Supplementary Fig. 4b is obtained as follows: first, the classical background of the conductance is subtracted using a 2$^{nd}$ order polynomial fitting. The conductance data is normalised by dividing the data by its standard deviation. Finally, the data is picked up at 0.025 T intervals to match the input size of the QGD. Supplementary Fig. 4c shows an output of the QGD from the conductance data.